\documentclass[aps,twocolumn,amsmath,amssymb,pre,floatfix]{revtex4-1}
\usepackage{amssymb}
\usepackage{amsmath}
\usepackage{graphicx}
\usepackage[table]{xcolor}
\usepackage{epsfig}
\usepackage[utf8]{inputenc}
\usepackage{dcolumn}
\usepackage{bm}
\usepackage{array}
\usepackage{multirow} 
\begin{document}

\title{First-principles calculations of Raman vibrational modes in the fingerprint region for connective tissue}

\author{E. T. Sato}

\author{H. Martinho}

\email{herculano.martinho@ufabc.edu.br}

\affiliation{Centro de Ciências Naturais e Humanas, Universidade Federal do ABC, Av. dos Estados 5001, Santo André-SP, 09210-580, Brazil\\}

\begin{abstract}

Vibrational spectroscopy has been widely employed to unravel physical-chemical properties of biological systems. Due to its high sensitivity to monitor real time "in situ" changes, Raman spectroscopy has been successfully employed, e.g., in biomedicine, metabolomics, and biomedical engineering. The grounds of interpretation of Raman spectra in these cases is the isolated macromolecules constituent vibrational assignment. Due to this, probe the anharmonic interactions or the mutual interactions among specific moieties/side chains to name but a few is a challenge. We present a complete vibrational modes calculation for connective tissue in the fingerprint region ($800-1800$ cm$^{-1}$) using first-principles Density Functional Theory. Our results indicated that important spectral features correlated to molecular characteristics have been ignored within the usual tissue spectral bands assignments. In particular, we found that the presence of confined water is the main responsible for the observed spectral complexity. Our calculations accounted for the inherent complexity of the spectral features in this region and useful spectral markers for biological processes were unambiguously identified.

\end{abstract}

\maketitle

The rapid, noninvasive and high spatial resolution capabilities of Raman spectroscopy technique have been employed to obtain biochemical and structural pieces of information of biological samples. Biomedicine\cite{butler2016using}, metabolomics \cite{santos2015rapid}, biomedical engineering\cite{e2011diagnosis,e2016optical} are examples of fields where this tool have been successfully used to acquire high-quality data. In particular, several optical-biopsy studies have shown that molecular interaction features in cells and tissues which cannot be accessed by conventional histopathology can be probed by this technique\cite{martinholivro}. Raman is of special interest due to their high sensitivity to detect  biochemical and molecular variations in tissues\cite{martinholivro}.

To a first approximation, spectrum of biological tissue is a convolution of isolated biological macromolecules (e.g., carbohydrates, proteins, lipids, deoxyribonucleic acid, ribonucleic acid) spectra. Hence, the tissue vibrational bands assignment is usually made based on isolated macromolecules assignment. There are several literature compilations (see, e.g., refs. \cite{movasaghi2007raman,movasaghi2008ftir,gelder2007,tuma2005} ) listing the vibrational bands of biological tissues constituents. These compilations eventually are used to perform a qualitative interpretation of the spectra. However, a large amount of relevant pieces of information are absent to one using this qualitative vibrational assignment. Anharmonic interactions which gives rise to coupling among harmonic vibrational modes is an example\cite{Sibert2015, Shikhovtseva201647}. Moreover, mutual interactions among specific moieties could be analyzed only in comparative basis. The usage of computational simulations of  small specific parts and short time intervals of the macromolecule is another method to understand this interactions. These approaches usually obscure relevant physical-chemical data from the environment. At scale of real biology they are in fact only a small part of the overall picture\cite{saunders2013coarse}.  

Computer simulations could be a suitable tool to interpret experimental data aiming understand biochemical changes translating structural changes that lead to macroscopic biological processes. Vibrational spectra of macromolecules and tissues are a important class of experimental data addressing this issue. The atomistic models based on quantum mechanical calculations have better prediction of the materials properties. However, due to its inherent complexity, atomistic modeling of biological systems are still in the early stages. In a previous\cite{sato} we presented a computational model for skin (STmod). The model consisted of a collagen peptide cutout including confined water submitted to periodic boundary conditions. The model was able to successfully explain important experimental structural and general biochemical trends of normal and inflammatory tissues.

\begin{figure}[tbh!]
	\includegraphics[width=8.0cm]{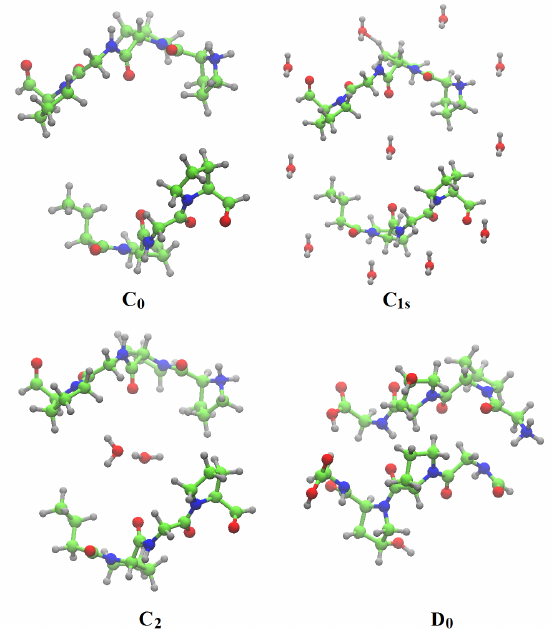}
	\caption{Unit cells for $C_0$, $C_{1s}$,$C_{2}$, and $D_{0}$ unit cells. Structural pieces of information could be found on Table I of ref. \cite{sato}.}\label{clbm_clean}
\end{figure}

In the present work a detailed vibrational modes assignment of a connective tissue based on the STmod is presented. To the best of our knowledge this is the first report on literature concerning complete vibrational assignment of a tissue. The vibrational calculations were performed on $C_{n}$ ($n-8$), $C_{1s}$, $D_0$, and $D_1$ unit cells of STmod. The numeric subscript indicates the number of water molecules inside the unit cell. The "s" subscript related to the presence of external water solvating the $C_1$ model.  Starting from a hydrated collagen peptide each unit cel was obtained and calculations performed on periodic boundary conditions. More details concerning the obtainment of these structures and previous characterizations as well could be found in ref. \cite{sato}. Figure \ref{clbm_clean} shown the unit cell for $C_{0}$, $C_1s$, $C_2$, and $D_0$ structures.

\begin{figure}[tbh!]
	\includegraphics[width=8.0cm]{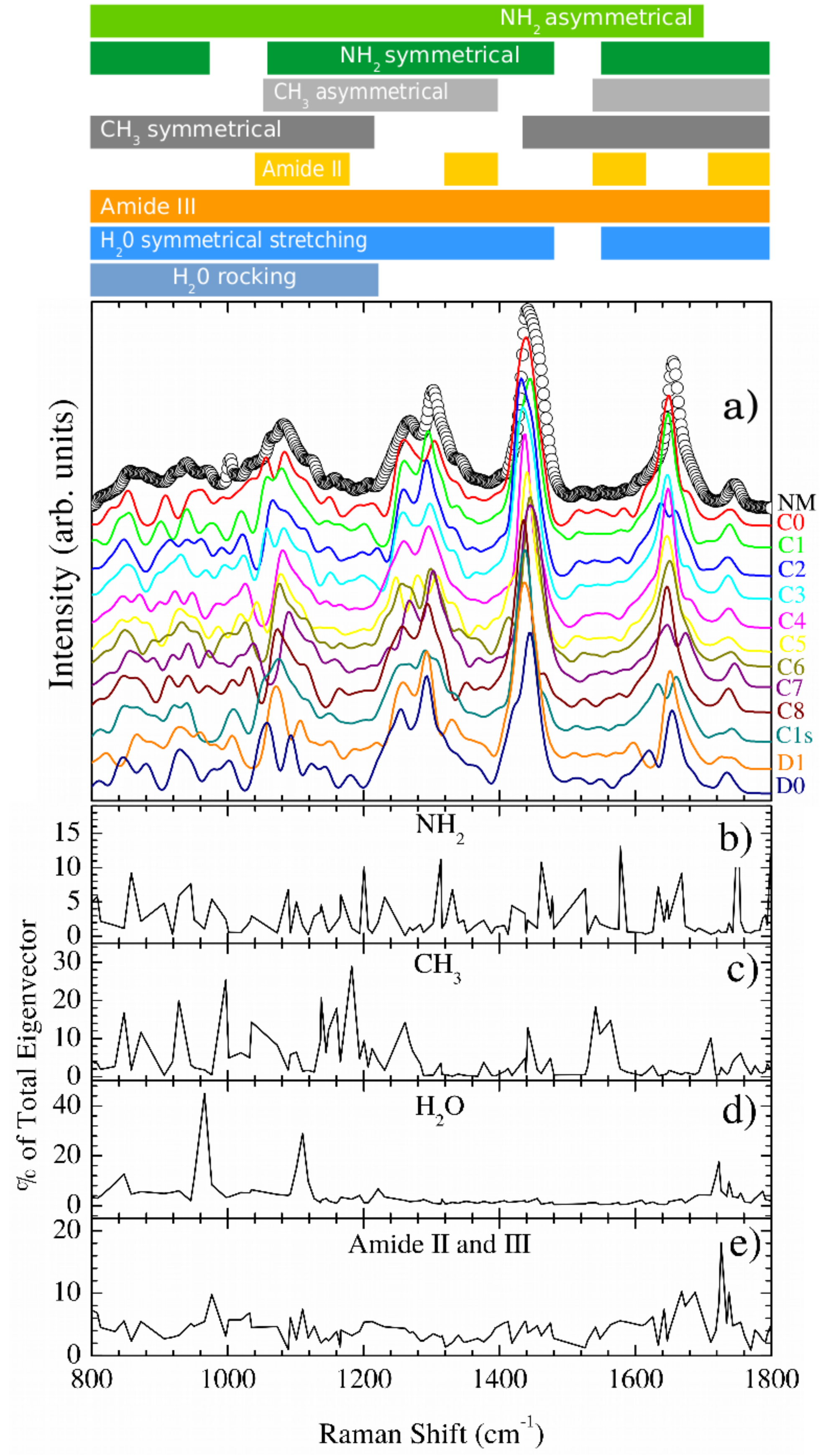}
	\caption{a) Experimental Raman spectra for normal oral mucosa tissue (NM) compared to $C_{n}$ ($n=0-8)$, $C_{1s}$, $D_0$, and $D_1$ STmod models in the fingerprint region.The wavenumber-axis projection of amine (b), methyl (c), water (d), and Amide II and III partial eigenvectors contributions are also shown. The top scheme represents the qualitative additional bands assignment based on Table \ref{table9}.}
	\label{Raman1}
\end{figure}

Density Functional Theory (DFT)\cite{hohenberg1964inhomogeneous,kohn1965self} was used in order to obtain the equilibrium geometries and harmonic frequencies. The simulations were implemented in the CPMD program\cite{cpmd} using the BLYP functional\cite{lee1988development} augmented with dispersion corrections for the proper description of van der Waals interactions\cite{von2005performance,lin2007library}. The cutoff energy was considered up to $100$ Ry. The wave functions were optimized and then the vibrational modes were obtained using the Hessian matrix. Finally the linear response for the values of polarization and polar tensors of each atom in the system was calculated to evaluate the eigenvectors of each vibrational mode.The Raman-active modes were obtained from the atomic polar tensors for each atom in the system and the corresponding eigenvectors of Hessian\cite{leach2001}. Harmonic frequencies were compared to experimental Raman data of normal(NM) oral mucosa tissue (see ref. \cite{e2011diagnosis} for experimental details). The spectra were simulated as a convolution of Gaussian lineshape peaks centered on the calculated frequencies using the Fityk \cite{fityk} program . The linewidth was chosen to be $20$ cm$^{-1}$. Figure \ref{Raman1}a) presents the results for fingerprint region. The experimental Raman spectrum of NM is also shown.

From direct inspection, we found that $C_2$ and $C_3$ models presented a large set of represented bands (thirteen). The worst model was $C_{7}$ one, which has only four peaks according to the experimental results, followed by $C_{0}$, consistent with only five peaks. The $C_3$ model presented a set of negative high frequency ($\gtrsim 940$ cm$^{-1}$) modes indicating some degree of mechanical instability. Thus, we concluded that $C_2$ is the suitable model to represent the connective tissue from the vibrational modes point of view.

\begin{table*}
	\caption{\label{table9}Connective tissue Raman bands assignment based in $C_{2}$ model. Additional vibrations compared to usual assignment in the fingerprint region are shown in \textcolor{red}{red}.  It is  important to notice that torsion and stretching at least one of the rings, C-C, CH, CN, NH, C=O, CH$_{3}$, OH stretching,  CH$_{2}$ symmetrical and asymmetrical stretching, Amide III, and H$_{2}$O scissoring vibrations appeared in all bands in some degree.}
	\begin{ruledtabular}
		\begin{tabular}{cm{14.5cm}}
			Experimental (cm$^{-1}$) & Assignment \\ 
			\hline
			$800-880$ & \textcolor{red}{CH$_{3}$ symmetrical stretching}; C-C stretching; and C-C-H deformation usually of proline, hydroxyproline, and collagen backbone \\ 
			\hline
			$880-980$ & \textcolor{red}{CH$_{3}$ and confined H$_{2}$O symmetrical stretching; C=O wagging; CH$_{2}$ twisting; CH$_{2}$, CH$_{3}$ and confined H$_{2}$O rocking};  C-C stretching; and C-C wagging usually of proline, hydroxyproline, and collagen backbone  \\ 
			\hline
			$980-1,050$ & \textcolor{red}{CH$_{3}$ symmetrical and asymmetrical stretching; CH and CH$_{2}$ twisting; CH$_{2}$ and CH$_{3}$ rocking; CH$_{2}$ scissoring}, C-C aromatic ring stretching usually of proline and collagen\\ 
			\hline
			$1,050-1,150$ & \textcolor{red}{CH$_{3}$ and NH$_{2}$ symmetrical and  asymmetrical stretching; CH$_{2}$ wagging; CH$_{2}$ twisting; CH$_{3}$ rocking}, C-C stretching; C-O stretching; C-N stretching vibration; usually of collagen, several bands of moderate intensity, belonging to Amide III and other groups (proteins)  \\
			\hline
			$1,150-1,230$ & \textcolor{red}{CH$_{3}$ and NH$_{2}$ symmetrical and asymmetrical stretching; CH wagging; CH$_{2}$ twisting; CH$_{2}$, CH$_{3}$ rocking}; C-C aromatic; CH$_{2}$ wagging vibrations from proline side chains; C-C/C-N stretching (proteins); several bands of moderate intensity, belonging to Amide III and other groups (proteins), differences in collagen content, hydroxyproline, proteins, including collagen I  \\
			\hline
			$1,230-1,280$ & \textcolor{red}{CH$_{3}$ asymmetrical stretching; CH wagging; CH$_{2}$ twisting}; CH$_{2}$ wagging; C-N stretching; C-N in plane stretching; CH$_{\alpha\prime}$rocking; second Amide; several bands of moderate intensity, belonging to Amide III and other groups (proteins), differences in collagen content, protein band \\
			\hline
			$1,280-1,320$ & \textcolor{red}{CH$_{3}$ asymmetrical stretching; CH wagging; CH$_{2}$ and CH$_{3}$ rocking}, CH$_{2}$; CH$_{3}$ twisting; CH$_{2}$ bending; CH$_{2}$ wagging; CH$_{2}$ deformation; several bands of moderate intensity, belonging to Amide III and other groups (proteins); differences in collagen content, bending mode of collagen  \\
			\hline
			$1,320-1,400$ & \textcolor{red}{NH$_{2}$ symmetrical and asymmetrical stretching; NH$_{2}$ rocking}, CH and CH$_{2}$ deformation; CH$_{2}$ bending; CH$_{3}$; CH$_{2}$ twisting; CH rocking; C=O symmetric stretch; CH$_{2}$ wagging in collagen; CH$_{3}$ and CH$_{2}$ deforming modes of collagen, several bands of moderate intensity, belonging to Amide III and other groups (proteins)  \\
			\hline
			$1,400-1,480$ & \textcolor{red}{CH$_{3}$ symmetrical stretching; NH$_{2}$ symmetrical and  asymmetrical stretching; CH$_{2}$ twisting; CH$_{2}$, CH$_{3}$ and NH$_{2}$ rocking}; CH$_{2}$ stretching/CH$_{3}$ asymmetric deformation; CH$_{2}$ wagging; CH, CH$_{2}$, CH$_{3}$ deformation; CH$_{2}$ scissoring; C-H and CH$_{2}$ bending mode of proteins, collagen\\
			\hline
			$1,480-1,540$ & \textcolor{red}{NH$_{2}$ asymmetrical stretching; NH$_{2}$ rocking}; C-N stretching vibration C=C stretching in benzenoid ring, Amide II, collagen  \\
			\hline
			$1,540-1,620$ & \textcolor{red}{CH$_{3}$ and confined H$_{2}$O symmetrical stretching; CH$_{3}$ and NH$_{2}$ asymmetrical stretching; CH, CH$_{2}$ and OH wagging; CH$_{2}$ and NH$_{2}$ twisting; CH$_{2}$ and NH$_{2}$ rocking}; C-C stretching; Amide I; Amide II  \\
			\hline
			$1,620-1,700$ & \textcolor{red}{CH$_{3}$, NH$_{2}$ and H$_{2}$O symmetrical stretching; CH$_{3}$ and NH$_{2}$ asymmetrical stretching; CH and CH$_{2}$ wagging; CH$_{2}$ and NH$_{2}$ twisting; CH$_{2}$ and CH$_{3}$ rocking; CH and CH$_{2}$ scissoring}; C=C stretching band; C=O stretching of collagen, Amide I, differences in collagen content, intermolecular bending mode of water, proteins, including collagen I  \\
			\hline
			$1,700-1,800$ & \textcolor{red}{CH$_{3}$, NH$_{2}$ and H$_{2}$O symmetrical stretching; CH$_{3}$ asymmetrical stretching; Amide II; CH$_{2}$ and OH wagging; CH$_{2}$ and NH$_{2}$ twisting; CH$_{2}$, CH$_{3}$ and NH$_{2}$ rocking; CH$_{2}$ scissoring}; Amide I; C=O stretching vibrations  \\
				\end{tabular} 
	\end{ruledtabular}
\end{table*}

Connective tissue vibrational modes based on $C_2$ model are shown in Table \ref{table9}. The comparison between the present work and literature bands assignment are also shown. We notice that our results presented extra pieces of information. The most striking feature is the activation of methyl, methylene, and amine side chains vibrations along the fingerprint spectral window.  Figure \ref{Raman1}b)-\ref{Raman1}e) shows the percentage contribution to the total eigenvectors projected on the wavelength axis for some of these vibrations. Amine vibrations (Fig. \ref{Raman1} b) are present on almost the entire region weighting around $10\%$ of contribution. It is possible to observe that methyl groups contributions (Fig. \ref{Raman1}c) are $\gtrsim 10\%$ below $1300$ cm$^{-1}$ and around $1450$, $1550$, and $1700$ cm$^{-1}$. The side chains are key factors determining the properties and reactivity of molecules. Thus, one expect that molecular transformations under, e.g., pathological processes, will display overall changes in the fingerprint vibrational region. It is important to notice that our calculations indicated that the side chains vibrational activation occurs only in the presence of confined water. The anhydrous C$_0$ model did not display this characteristic. The water dimer itself presents spectral features around $900$ and $1600-1750$   cm$^{-1}$ (Fig. \ref{Raman1} d) which usually are not described. Amide II and III vibrations are also present on the overall spectra weighting around $5\%$ (Fig. \ref{Raman1}e).

In fact, experimentally observed spectral changes in fingerprint region for tissues have been qualitatively reported to correlated to water content (see,e.g., refs. \cite{pereira2015rm1,lopes2017vivo,barroso2015discrimination,da2008role}). Elderly and diabetis\cite{pereira2015rm1,lopes2017vivo}, oral cancer\cite{barroso2015discrimination}, cervical cancer\cite{da2008role} to name but a few are examples of physiological situations where spectral complexity emerges beyond the isolated molecule vibrational bands assignment. Protein-water  interactions  are  known  to  play  a  critical  role in the function of several biological systems and macromolecules including collagen in tissues\cite{fathima2010structure}. Small  changes  in  structure  and  dynamical behavior of water molecules at the peptide-water interface can effectively change both the structure and dynamics of the protein function\cite{lima2012anharmonic}. Our model indicates that the main source of this complexity is the presence of confined water enabling distant and isolated side chains coupling. A large set of wagging, scissoring, twisting and rocking vibrations of side chains usually assigned in the high-wavenumber region ($\gtrsim 2000$ cm$^{-1}$) appeared in the fingerprint region damped to usually assigned vibrations (see Table \ref{table9}). Coexistence of symmetrical and asymmetric stretchings are also present in a more complicated fashion that go beyond the protein, collagen, proline and hydroxyproline usual bands assignment. Interestingly, the region $880-940$ cm$^{-1}$ appeared to retain information about the confined water content (Fig. \ref{Raman1} c). In fact, the contributions to the eigenvectors in this region goes from $30\%$ confined water and $70\%$ of {CH$_{3}$,CH,C-C,C-C-H} vibrations.

From the qualitative and quantitative pieces of information generated by our first principle calculations one can found useful spectral markers which could be correlated to biological process of interest. We will comment two examples. 

\textit{Confined Water}. Since the $800-880$ region is dominated by {C,H} vibrations (Table \ref{table9}) the difference between the integrated areas of these two regions will be a suitable qualitative quantifier for the confined water content,

\begin{equation}
[H_{2}O]_{confined}\varpropto I_{880-940}-I_{800-880}
\end{equation}
\noindent a key parameter for describing important processes, as just commented in the previous discussion.

\textit{Methylation}. The $800-880$ cm$^{-1}$ region could also be used to probe the protein methylation process. Protein methylation is the process through which methyl groups are added to proteins under the action of specific enzymes, the methyltransferases. Usually it occurs on nitrogen atoms in N-terminals and cannot be reversed creating new amino acid residues.\cite{clarke1993protein}. The above-cited spectral region does not present amine nor hydroxil contributions being exclusive of methyl and protein backbone vibrations. Thus computing their ratio to N-H stretching appearing in the high-wavenumber region \cite{e2011diagnosis},

\begin{equation}
Protein_{CH_{3}}=I_{800-880}/I_{N-H}
\end{equation}

\noindent will give a useful protein methyl quantifier. Usually the methyl band in the $\sim 2940$ cm$^{-1}$ high-wavenumber region is considered to evaluate the methylation. However, it includes contributions from lipids and proteins\cite{e2011diagnosis}.  Protein methylation modulates cellular and biological processes including transcription, RNA processing, protein interactions and protein dynamics\cite{AFJEHISADAT201312}. Methyl-binding protein domains and improved antibodies with broad specificity for methylated proteins are being used to characterize the so-called \textit{protein methylome}. They also have the potential to be used in high-throughput assays for inhibitor screens and drug development\cite{emergingprotein}.

In summary, our vibrational modes calculation for connective tissue in the fingerprint region indicated that important spectral features correlated to molecular characteristics have been ignored within the usual tissue spectral bands assignments. Our results indicated that the presence of confined water is the the main responsible for the observed spectral complexity being a factor that cannot be ignored. The inherent complexity of the spectral features in this region could be rationalized by our calculations and useful spectral markers for biological processes could be identified.

\textbf{Acknowledgements} The authors would like to thank the Brazilian agencies Conselho Nacional de Desenvolvimento Científico e Tecnológico (CNPq - 311146/2015-5) and Fundação de Amparo à Pesquisa do Estado de São Paulo (FAPESP - 2011/19924-2) for the financial support. We would also thank the computational resources provided by Centro Nacional de Processamento de Alto Desempenho em São Paulo (CENAPAD-UNICAMP) and Sistema de Computação Petaflópica (Tier 0) (Santos Dumont-LNCC) under  Sistema Nacional de Processamento de Alto Desempenho (SINAPAD) of the  Ministério da Ciência, Tecnologia e Inovação (MCTI).

\end{document}